\documentclass{vgtc}                          

\graphicspath{{figures/}{pictures/}{images/}{./}} 

\usepackage{times}                     

\usepackage{tabu}                      
\usepackage{booktabs}                  
\usepackage{lipsum}                    
\usepackage{mwe}                       

\usepackage{mathptmx}                  
\usepackage{subfiles}

\onlineid{0}

\vgtccategory{Research}

\vgtcinsertpkg

\title{Discussing Your Needs in VR: A Novel Approach through Persona-based Stakeholder Role-Playing}

\author{Yi Wang\thanks{e-mail: xve@deakin.edu.au}\\ %
        \parbox{1.7in}{\scriptsize \centering School of Information Technology,  \\ Deakin University}%
\and Zhengxin Zhang\thanks{e-mail: robinzhang2015@iCloud.com}\\ %
     \parbox{1.7in}{\scriptsize \centering School of Design,  \\ Inner Mongolia Normal University} %
\and Xiao Liu\thanks{e-mail: xiao.liu@deakin.edu.au}\\ %
     \parbox{1.7in}{\scriptsize \centering School of Information Technology,  \\ Deakin University}
\and Chetan Arora\thanks{e-mail: chetan.arora@monash.edu}\\ %
     \parbox{1.7in}{\scriptsize \centering Faculty of Information Technology,  \\ Monash University}
\and John Grundy\thanks{e-mail: john.grundy@monash.edu}\\ %
     \parbox{1.7in}{\scriptsize \centering Faculty of Information Technology,  \\ Monash University}
\and Thuong Hoang\thanks{e-mail: thuong.hoang@deakin.edu.au}\\ %
     \parbox{1.7in}{\scriptsize \centering School of Information Technology,  \\ Deakin University}
     }

\teaser{
  \centering
  \includegraphics[width=\linewidth]{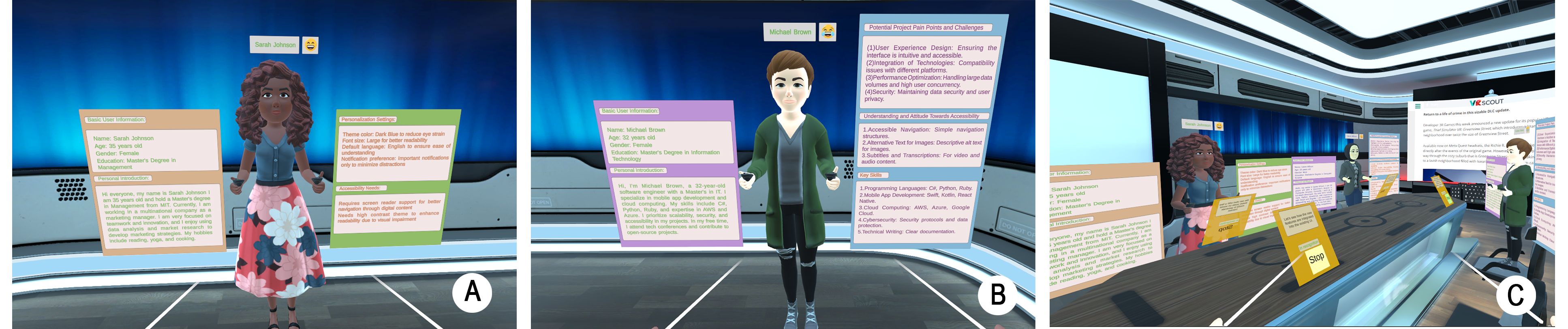}
  \caption{\textbf{A} and \textbf{B} represent personas generated from different types of speech-to-text content. \textbf{C} represents the virtual environment used for requirements discussion.}
  \label{fig:teaser}
}

\abstract{
In this study, we propose a novel approach that supports requirements discussions in virtual environments by automatically generating personas from real-time speech-to-text data. In our pilot experiment, 18 participants (14 from universities and 4 from IT companies) used the generated personas to discuss accessibility requirements within the virtual environment. Participants reported a relatively high level of satisfaction with the social presence and usability of the VR system. We also found that requirements discussions based on personas have a lower workload. Finally, we outline the main directions for future work.
} 


\begin{document}



\maketitle

\section{Introduction}

There has been an increasing focus on the use of virtual reality (VR) for requirements engineering (RE) activities, especially for the requirements elicitation or gathering phase \cite{Wang22}. Among these, VR provides significant value in social collaboration among stakeholders. For example, VR-based social collaboration has been shown to positively enhance the sense of participation among older adults in virtual environments \cite{kalantari2023using}. Meanwhile, recent work has explored the use of VR to support requirements engineering activities, such as requirements elicitation. Wang et al. \cite{Wang21} proposed a series of novel VR features, such as automatically generating 3D avatars, observing user behavior from a first-person perspective, and visual impairment simulators. However, this work did not include a formal user study or a fully functional system. Wang et al. \cite{Wang22} further introduced a preliminary system that supports real-time communication and interaction while capturing keywords from conversations. Although many VR platforms support remote meetings and discussions, these basic functionalities are insufficient for supporting requirements discussions in virtual environments. 


Personas as a mainstream method in RE, are frequently used in early-stage requirements discussions \cite{Karolita23}. To date, no work has explored the advantages of using automatically generated personas to support requirements discussions in virtual environments. Therefore, our motivation is to explore whether applying automatically generated personas to requirements discussions in virtual environments can effectively support practitioners in discussing requirements, particularly for multinational companies or software teams that prefer using VR as a communication medium.

To fill this gap, we propose a novel approach that integrates automatically generated personas within a virtual environment. Our VR system supports networked communications for multiple stakeholders, with real-time speech-to-text capabilities and a GPT-4-based text analysis model. We asked participants to discuss, within the virtual environment, whether a case website met accessibility requirements based on the automatically generated personas. The results showed that our system provided a higher social presence and system usability, and that the VR-based approach significantly reduced participants’ perceived workload.

\section{System Design}


Figure \ref{fig:teaser} C shows an overview of our VR-based requirements discussion system. Figure \ref{fig:teaser} A and B show the automatically generated user personas based on real-time discussion data. Prior to entering the VR environment, participants were able to select their preferred avatars. Thus, we used open-source avatars provided by Meta.





We enabled multi-stakeholder communication through real-time voice and synchronous avatar interactions. Photon was employed to enable low-latency audio transmission and networked synchronization. A 3D Web View was embedded in the scene, allowing participants to collaboratively discuss accessibility requirements based on predefined case websites. All participant audio was processed in real time via Azure cloud services, with audio and transcription using automatic speech-to-text. To maintain data clarity, the system stores each participant’s audio stream separately, avoiding overlap and preserving the integrity of individual contributions.




After the discussion, the transcription is submitted to a GPT-4–based analysis pipeline. The system supports four key prompts: (1). Generate personas from participants audio data, including demographic attributes. If certain details (e.g., names) are missing, the system can reasonably infer or fabricate them. (2). Generate personas from participants' audio data, incorporating demographic details, potential accessibility requirements, personalized settings, and inferred biographies. (3). Extract participants' attitudes, identifying their understanding of accessibility, project pain points, challenges, and relevant expertise. (4). Identify user needs and preferences, focusing on accessibility requirements and personal interests derived from audio data. Additionally, the system performs emotion recognition based on textual input from participants, visualized through emoji representations of emotional states.

\section{Experiment}

\begin{figure}[t]
 \centering 
 \includegraphics[width=0.61\linewidth]{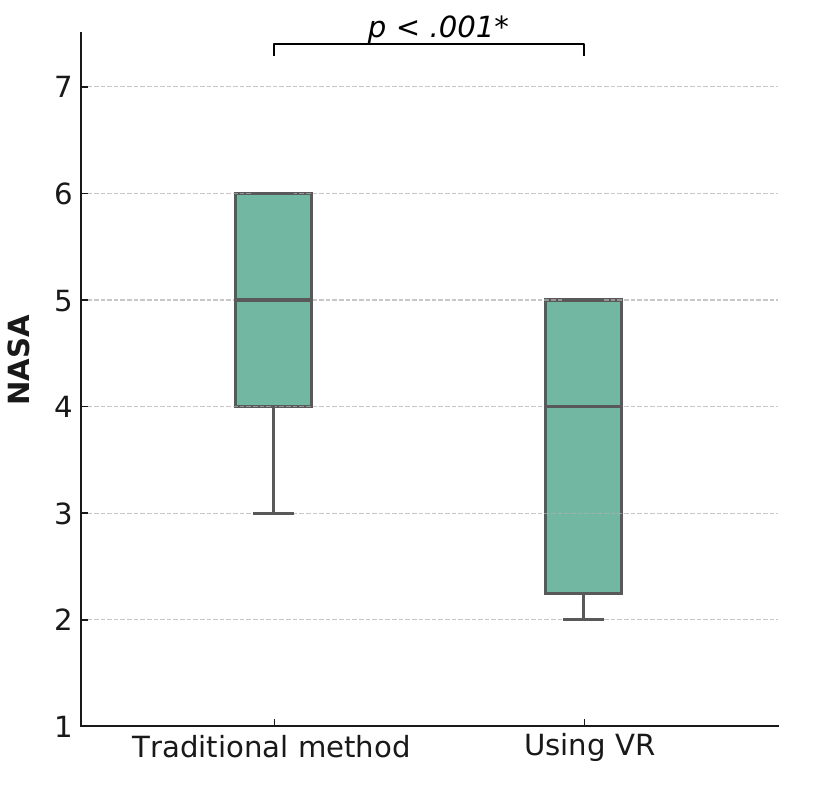}
 \caption{Box plot for results of workload questionnaires. }
 \label{fig:1_all}
 \end{figure}

We employed a ``within-subjects'' design for our experiment. The experiment included user needs discussion in two conditions: using automatically generated personas within a virtual meeting room (VR method), and using manually create personas within a physical meeting room (Traditional method).


\subsection{Participants}

We recruited 18 participants (10 males and 8 females) in seven pairs from both our university courses and IT companies, with their ages ranging from 21 to 29 years (\textit{M} = 25.9, \textit{SD} = 3.1). Among these, a total of 14 participants (8 males and 6 females, aged 21-24) were recruited from our university, all of whom were undergraduate and graduate students, from interaction design (\textit{N} = 7) and computer science (\textit{N} = 7). The remaining four participants were from IT companies and were primarily engaged in website design and development (3 males and 1 female, aged 25-29).

\subsection{Experiment Procedure}

Upon arrival, participants read an information sheet and signed a consent form. Participants were randomly assigned to either the traditional method or the VR-based method, and after completing the task, they were assigned to the other condition. When using the VR-based method, they were guided to wear the Meta Quest 3, explore the default Welcome Lobby, adjust the headset, launch the application, select an avatar, and enter the virtual scene. Basic interaction methods were introduced. When using the traditional method, they were asked to discuss accessibility requirements based on an existing website.

Participants first introduced themselves (e.g., age, hobbies, background) and then explored the website’s features, such as clickable elements, layout, text size, video player, and input fields. Participants described their accessibility requirements and challenges (e.g., lack of font size adjustment or poor navigation), potential solutions and constraints (e.g., technical or resource limitations). Next, participants generated personas and requirements based on recorded speech and continued the discussion using the personas and emotional cues (via emojis). Participants reviewed each other’s personas, discussed accessibility requirements, and collaboratively prioritized key requirements and necessary resources. Finally, after a short break, participants completed three 7-point  Likert Scale Questionnaires that included the social presence, system usability, and NASA-TLX. Social presence and system usability are only applicable to VR-based requirement discussion systems. The entire experiment lasted approximately 1 hour and 30 minutes.

\subsection{Results and Discussion}



Descriptive statistics indicated relatively high scores for the VR system in terms of social presence (\textit{Mean} = 5.21, \textit{SD} = 0.81) and system usability (\textit{Mean} = 5.19, \textit{SD} = 0.78). Therefore, participants experienced a strong sense of co-presence when using the VR system and generally perceived the system as usable, such as reporting clear interaction logic and low task burden. However, a key limitation of this study is that we did not evaluate the system against existing VR meeting platforms or persona-based systems. In future work, we plan to compare our VR system with existing systems.

Figure \ref{fig:1_all} shows the results of the workload. We observed a significant difference between the traditional method and the VR method (\textit{p} $<$ .001). Descriptive statistics indicated a lower score on workload for using VR (\textit{Mean} = 3.57, \textit{SD} = 0.85). Overall, discussing requirements using automatically generated personas within the VR system required less workload than face-to-face discussions with manually created personas. This advantage may influence practitioners’ attitudes toward the use of personas, particularly among VR practitioners. Thus, compared to manually created personas, automatically generated personas in VR may offer previously unexplored benefits. For example, whether they provide advantages over traditional approaches or LLM-based persona generation systems implemented on web platforms. However, a key limitation at present is the level of acceptance of VR as a tool among software practitioners.

\section{Conclusion}

In this paper, we introduce a VR system that uses automatically generated personas to facilitate discussions about requirements in a virtual environment. These personas are created based on the audio data converted to text from the participants. We conducted a preliminary study to evaluate the user experience of our VR-based requirements discussion system. A total of 18 participants were recruited, including 14 students and 4 industry practitioners. The results showed relatively high scores for the VR system in terms of social presence and system usability. The results also indicated that the use of personas to discuss requirements reduces workload. 



\bibliographystyle{abbrv-doi}

\bibliography{template}
\end{document}